# Stationary photon-atom entanglement and flow equation

Yuichi Itto[1] and Sumiyoshi Abe[1,2]

[1]*Department of Physical Engineering, Mie University, Mie 514-8507, Japan*

[2]*Institut Supérieur des Matériaux et Mécaniques Avancés,*

*44 F. A. Bartholdi, 72000 Le Mans, France*

**Abstract** Stationary photon-atom entanglement is discussed by applying the method of flow equation to the Jaynes-Cummings model. A nonlocal continuous unitary transformation is explicitly constructed and the associated positive operator-valued measures for the photons and atom are obtained. Then, flow of the entanglement entropy is analyzed. A comment is also made on implementing the unitary operation in the method of flow equation. This method may offer a new strategy for quantum-state engineering.

PACS numbers: 03.65.Ud, 42.50.-p, 03.65.Db

## I. INTRODUCTION

A recent major breakthrough in experimental quantum information would be the realization of quantum teleportation between photons and atoms [1]. A crucial step in the experiment is to generate entanglement between polarization of light and caesium atoms by the use of the quantum-mechanical Faraday rotation. This development is of fundamental importance toward fault-tolerant quantum computation and quantum communication.

Importance of photon-atom entanglement sheds fresh light on traditional quantum optical systems. Actually, there are some discussions about photon-atom entanglement in the literature. They are, however, concerned with nonstationary states with rapid temporal oscillations of entanglement [2,3], which may not be so useful for quantum information processing. Thus, a point is how to generate stationary entangled states in a photon-atom system. *Mathematically*, this may be nothing else than diagonalizing the system Hamiltonian. *Experimentally*, however, given an initial state, it is hard to realize the quantum operation that can immediately transform such a state of the coupled system to the exact stationary state, in general. Therefore, in reality, it is necessary to inquire into a series of “trial” operations in order to approach the stationary state. Then, the question is what a desirable strategy for it is.

To answer to the above question, here we examine the method of flow equation introduced by Wegner [4,5]. This theory uses a continuous unitary transformation depending on a single parameter, $l$. The Hamiltonian tends to become diagonalized in the limit, $l \to \infty$. It is our opinion that Wegner's flow equation has physical meanings more than just a procedure of diagonalizing the Hamiltonian. In fact, it may offer the *optimal* strategy for quantum-state engineering [6]. In other words, *if the trial operations are efficient, they are almost on a trajectory generated by the flow equation*. Due to this observation, it seems meaningful to examine the flow equation approach for quantum optical systems of photons and atoms.

In this paper, we therefore consider the flow equation approach to establishing stationary entanglement between light and matter. As a simple but nontrivial example, we study the celebrated Jaynes-Cummings model [7,8] and perform its complete flow-equation analysis. Then, we shall calculate the positive operator-valued measures (POVMs) for the photons and atom, evaluate the degree of stationarity, and analyze flow of the entanglement entropy. We also present a procedure for implementing the unitary operation appearing in the method of flow equation for the model.

## II. BASICS OF FLOW EQUATION

Let us recapitulate the method of flow equation. Consider the continuous sequence of unitary transformations parametrized by $l$: $U(l)$. $l$ is referred to as the flow parameter, which has dimension $(\text{energy})^{-2}$. Application of $U(l)$ to a Hamiltonian $H$ yields $H(l) = U(l) H U^{\dagger}(l)$, which obeys the following differential equation:

$$\frac{dH(l)}{dl} = [\eta(l), H(l)] \tag{1}$$

with the “initial” condition $H(0) = H$. $\eta(l) \equiv \left[dU(l)/dl\right] U^{\dagger}(l)$ is an anti-Hermitian operator. Wegner’s proposal for the choice of this operator is

$$\eta(l) = [H^{\mathrm{d}}(l), H(l)], \tag{2}$$

where $H^{\mathrm{d}}(l)$ stands for the diagonal part of $H(l)$. With this choice, it can be shown [4,5] that the off-diagonal matrix elements of $H(l)$ exponentially decay in $l$ if $H^{\mathrm{d}}(l)$ is nondegenerate. Eq. (1) defines the flow equations of the coefficients contained in $H(l)$. This is in analogy with the concept of renormalization group.

## III. FLOW EQUATION OF THE JAYNES-CUMMINGS MODEL

Let us apply the method of flow equation to the Jaynes-Cummings model [7,8]. This model describes quantum dynamics of the interaction between a single-mode radiation field and a two-level atom in the rotating wave approximation. The Hamiltonian reads

$$H = \frac{\omega_0}{2}\sigma_3 + \omega a^{\dagger} a + \lambda(\sigma_+ a + \sigma_- a^{\dagger}). \tag{3}$$

The symbols appearing in this equation are defined as follows. $\sigma_3 = |e\rangle\langle e| - |g\rangle\langle g|$, $\sigma_+ = |e\rangle\langle g|$, $\sigma_- = |g\rangle\langle e|$ are the operators of the two-level atom with the normalized ground and excited states, $|g\rangle$ and $|e\rangle$. $a^\dagger$ and $a$ are the creation and annihilation operators of the photon satisfying the algebra: $[a, a^\dagger] = 1$, $[a, a] = [a^\dagger, a^\dagger] = 0$. $\omega_0$, $\omega$, and $\lambda$ are the frequency of transition between the ground and excited states of the atom, the frequency of the single-mode photon, and the real coupling constant between the photon and the atom, respectively. h is set equal to unity for the sake of simplicity. In the present work, we limit our analysis to the off-resonant case. Therefore, the detuning factor, $\Delta \equiv \omega_0 - \omega$, is nonvanishing, and here it is assumed to be positive:

$$\Delta = \omega_0 - \omega > 0. \tag{4}$$

Although this is a highly simplified model of the coupled system of light and matter, remarkably it can experimentally be realized to generate single-atom maser or micromaser in a cavity [9]. Since the model is exactly solvable, it allows to analytically discuss a variety of phenomena such as the Rabi oscillation and quantum collapses/revivals.

For the Jaynes-Cummings Hamiltonian in Eq. (3), we have constructed the following operator of the continuous unitary transformation:

$$U(l)=\sum_{n=0}^{\infty}\left[\alpha_n(l)\sigma_+\sigma_-|n\rangle\langle n|+\beta_{n-1}(l)\sigma_-\sigma_+|n\rangle\langle n|\right.$$

$$\left.+\gamma_n(l)\sigma_+|n\rangle\langle n+1|+\delta_n(l)\sigma_-|n+1\rangle\langle n|\right]. \qquad (5)$$

Here, $\beta_{-1}(l)=e^{i\chi(l)}$ [which turns out to be unity, later: see Eq. (27)] and $|n\rangle=(n!)^{-1/2}(a^\dagger)^n|0\rangle$ denotes the $n$-photon state with the normalized vacuum state $|0\rangle$, i.e., $a|0\rangle=0$. The unitarity condition, $U(l)\,U^\dagger(l)=U^\dagger(l)\,U(l)=I$, leads to the following set of equations ($n=0,1,2,\cdots$):

$$\left|\alpha_n(l)\right|^2+\left|\gamma_n(l)\right|^2=\left|\beta_n(l)\right|^2+\left|\delta_n(l)\right|^2=1,$$

$$\left|\alpha_n(l)\right|^2+\left|\delta_n(l)\right|^2=\left|\beta_n(l)\right|^2+\left|\gamma_n(l)\right|^2=1, \qquad (6)$$

$$\alpha_n(l)\delta_n^*(l)+\beta_n^*(l)\gamma_n(l)=0, \qquad \alpha_n(l)\gamma_n^*(l)+\beta_n^*(l)\delta_n(l)=0, \qquad (7)$$

from which, it follows that

$$\left|\alpha_n(l)\right|^2=\left|\beta_n(l)\right|^2, \qquad \left|\gamma_n(l)\right|^2=\left|\delta_n(l)\right|^2. \qquad (8)$$

The “initial” conditions are

$$\alpha_n(0) = \beta_n(0) = 1, \qquad \gamma_n(0) = \delta_n(0) = 0, \tag{9}$$

and $\beta_{-1}(0) = 1$. The transformed Hamiltonian is given by

$$H(l) = \sum_{n=0}^{\infty} \left[ A_n(l)\sigma_+\sigma_- |n\rangle\langle n| + B_{n-1}(l)\sigma_-\sigma_+ |n\rangle\langle n| \right.$$
$$\left. + C_n(l)\sigma_+ |n\rangle\langle n+1| + C_n^*(l)\sigma_- |n+1\rangle\langle n| \right], \tag{10}$$

where

$$A_n(l) = \frac{\omega_0}{2}\left[ |\alpha_n(l)|^2 - |\gamma_n(l)|^2 \right] + \omega\left[ n + |\gamma_n(l)|^2 \right]$$
$$+ \lambda\sqrt{n+1}\left[ \alpha_n(l)\,\gamma_n^*(l) + \alpha_n^*(l)\gamma_n(l) \right], \tag{11}$$

$$B_n(l) = \frac{\omega_0}{2}\left[ |\delta_n(l)|^2 - |\beta_n(l)|^2 \right] + \omega\left[ n + |\beta_n(l)|^2 \right]$$
$$+ \lambda\sqrt{n+1}\left[ \beta_n(l)\delta_n^*(l) + \beta_n^*(l)\delta_n(l) \right], \tag{12}$$

$$C_n(l) = \frac{\omega_0}{2}\left[ \alpha_n(l)\delta_n^*(l) - \beta_n^*(l)\gamma_n(l) \right] + \omega\,\beta_n^*(l)\gamma_n(l)$$
$$+ \lambda\sqrt{n+1}\left[ \alpha_n(l)\beta_n^*(l) + \gamma_n(l)\delta_n^*(l) \right], \tag{13}$$

for $n = 0, 1, 2, \cdots$, and $B_{-1}(l) \equiv -\omega_0 / 2$, provided that Eqs. (6) and (7) have been used.

Note that both $A_n(l)$ and $B_n(l)$ are real. The "initial" conditions are

$$A_n(0) = \frac{\omega_0}{2} + \omega n, \tag{14}$$

$$B_n(0) = -\frac{\omega_0}{2} + \omega(n+1), \tag{15}$$

$$C_n(0) = \lambda\sqrt{n+1}. \tag{16}$$

The diagonal part in Eq. (10) is given by

$$H^{\mathrm{d}}(l) = \sum_{n=0}^{\infty}\left[A_n(l)\sigma_+\sigma_-|n\rangle\langle n| + B_{n-1}(l)\sigma_-\sigma_+|n\rangle\langle n|\right]. \tag{17}$$

Accordingly, the anti-Hermitian operator in Eq. (2) is calculated to be

$$\eta(l) = \sum_{n=0}^{\infty}\left[A_n(l) - B_n(l)\right]\left[C_n(l)\sigma_+|n\rangle\langle n+1| - C_n^*(l)\sigma_-|n+1\rangle\langle n|\right], \tag{18}$$

leading to

$$[\eta(l), H(l)] = \sum_{n=0}^{\infty} \left[A_n(l) - B_n(l)\right] \left\{2\left|C_n(l)\right|^2 \left(\sigma_+\sigma_- |n\rangle\langle n| - \sigma_-\sigma_+ |n+1\rangle\langle n+1|\right)\right.$$

$$\left. -\left[A_n(l) - B_n(l)\right]\left[C_n(l)\sigma_+ |n\rangle\langle n+1| + C_n^*(l)\sigma_- |n+1\rangle\langle n|\right]\right\}. \qquad (19)$$

Therefore, Eq. (1) yields the following set of flow equations ($n = 0, 1, 2, \cdots$):

$$\frac{d A_n(l)}{d l} = 2\left[A_n(l) - B_n(l)\right]\left|C_n(l)\right|^2, \qquad (20)$$

$$\frac{d B_n(l)}{d l} = -2\left[A_n(l) - B_n(l)\right]\left|C_n(l)\right|^2, \qquad (21)$$

$$\frac{d C_n(l)}{d l} = -\left[A_n(l) - B_n(l)\right]^2 C_n(l). \qquad (22)$$

From Eq. (22), it is obvious that the off-diagonal matrix elements in fact decay as $l$ grows [note that $A_n(l) \neq B_n(l)$ because of the off-resonant condition in Eq. (4)]. From Eqs. (16) and (22) as well as the realities of $A_n(l)$ and $B_n(l)$, $C_n(l)$ also turns out to be real. The exact solutions to these equations under the “initial” conditions in Eqs. (14)-(16) are

$$A_n(l) = \omega\left(n + \frac{1}{2}\right) + \frac{\Delta}{2} L_n(l), \qquad (23)$$

$$B_n(l) = \omega\left(n + \frac{1}{2}\right) - \frac{\Delta}{2} L_n(l), \tag{24}$$

$$C_n(l) = \frac{\Omega_n}{2}\sqrt{1 - \frac{\Delta^2}{\Omega_n^2} L_n^2(l)}\,, \tag{25}$$

for $n = 0, 1, 2, \cdots$, where $\Omega_n = \sqrt{\Delta^2 + 4\lambda^2 (n+1)}$ is the Rabi frequency, and $L_n(l)$ is given by

$$L_n(l) = \frac{1}{\sqrt{\frac{\Delta^2}{\Omega_n^2} + \left(1 - \frac{\Delta^2}{\Omega_n^2}\right)\exp\left(-2\Omega_n^2 l\right)}}\,. \tag{26}$$

We note that without losing generality $A_n(l)$ is taken to be larger than $B_n(l)$. (If $\Delta$ were taken negative, then the case $A_n(l) < B_n(l)$ could be realized. This point does not affect the subsequent discussion.)

Furthermore, the equalities, $\eta(l) = [dU(l)/dl]\,U^{\dagger}(l)$ and $dU(l)/dl = \eta(l)U(l)$, together with Eqs. (6)-(9) and (18), lead to the fact that $\alpha_n(l)$, $\beta_n(l)$, $\gamma_n(l)$, and $\delta_n(l)$ are all real and satisfy the relations

$$\alpha_n(l) = \beta_n(l), \qquad \gamma_n(l) = -\delta_n(l), \qquad \beta_{-1}(l) = 1. \tag{27}$$

From these relations as well as Eqs. (6)-(8) and (27), the coefficients appearing in the unitary operator in Eq. (5) are found to be given as follows ($n = 0, 1, 2, \cdots$):

$$\alpha_n(l) = \beta_n(l) = \sqrt{\frac{1 + \tilde{L}_n(l)}{2}}, \tag{28}$$

$$\gamma_n(l) = -\delta_n(l) = \sqrt{\frac{1 - \tilde{L}_n(l)}{2}}, \tag{29}$$

where

$$\tilde{L}_n(l) = \frac{\Delta^2}{\Omega_n^2} L_n(l) + \sqrt{\left(1 - \frac{\Delta^2}{\Omega_n^2}\right)\left(1 - \frac{\Delta^2}{\Omega_n^2} L_n^2(l)\right)}. \tag{30}$$

Recalling the off-resonant condition in Eq. (4), Eqs. (23)-(25) respectively converge, in the limit $l \to \infty$, to

$$A_n(\infty) = \omega\left(n + \frac{1}{2}\right) + \frac{\Omega_n}{2} \equiv E_+(n), \tag{31}$$

$$B_n(\infty) = \omega\left(n + \frac{1}{2}\right) - \frac{\Omega_n}{2} \equiv E_-(n), \tag{32}$$

$$C_n(\infty) = 0, \tag{33}$$

with the energy eigenvalues, $E_{\pm}(n)$, corresponding to $\left|E_{+}\right\rangle = U^{\dagger}(\infty)|e\rangle|n\rangle$ and $\left|E_{-}\right\rangle = U^{\dagger}(\infty)|g\rangle|n+1\rangle$. Eq. (33) explicitly shows that the Hamiltonian is in fact exactly diagonalizable.

Closing this section, we point out that although the present flow equation approach assumes the off-resonant case as in Eq. (4), the limiting quantities, $A_n(\infty)$ and $B_n(\infty)$ (which are the energy eigenvalues in the states, $|e\rangle|n\rangle$ and $|g\rangle|n+1\rangle$, respectively) converge in the resonant limit to

$$A_n(\infty) \equiv E_{+}(n) \xrightarrow{\Delta\to+0} \omega\left(n+\frac{1}{2}\right)+\lambda\sqrt{n+1}, \tag{34}$$

$$B_n(\infty) \equiv E_{-}(n) \xrightarrow{\Delta\to+0} \omega\left(n+\frac{1}{2}\right)-\lambda\sqrt{n+1}, \tag{35}$$

which are the correct results obtained by other methods [7,8].

## IV. POSITIVE OPERATOR-VALUED MEASURES AND SUBSYSTEMS

In this section, we discuss the effects of the unitary operation by $U(l)$ on the subsystems, i.e., the photons and atom. We shall explicitly calculate the positive operator-valued measures (POVMs).

Consider $|\Psi\rangle = |e\rangle|n\rangle$ with $n \neq 0$, which is a natural “initial” state to be prepared since in a realistic experimental situation using a cavity the photons and atom may initially be uncorrelated each other. The corresponding density matrix of the total system reads

$$\rho_{\mathrm{total}} = |e\rangle\langle e| \otimes |n\rangle\langle n|. \tag{36}$$

Recalling the operation of $U(l)$ on the Hamiltonian, $H(l) = U(l) H U^{\dagger}(l)$, we see the corresponding operation on the state to be

$$|\Psi(l)\rangle = U^{\dagger}(l)|e\rangle|n\rangle. \tag{37}$$

Therefore, the above density matrix transforms as

$$\rho_{\mathrm{total}} \to \rho_{\mathrm{total}}(l) = U^{\dagger}(l)\rho_{\mathrm{total}} U(l). \tag{38}$$

The reduced density matrices of the subsystems are given by $\rho_{\mathrm{photon}}(l) = \mathrm{Tr}_{\mathrm{atom}}\, \rho_{\mathrm{total}}(l)$ and $\rho_{\mathrm{atom}}(l) = \mathrm{Tr}_{\mathrm{photon}}\, \rho_{\mathrm{total}}(l)$. The form in Eq. (36) allows us to obtain the following Kraus representations [10]:

$$\rho_{\mathrm{photon}}(l) = \sum_{i=g,e} A_i(l)|n\rangle\langle n|A_i^{\dagger}(l), \tag{39}$$

$$\rho_{\text{atom}}(l) = \sum_{m=0}^{\infty} B_m(l)|e\rangle\langle e|B_m^\dagger(l). \tag{40}$$

Here, $\{A_i(l)\}_{i=g,e}$ and $\{B_m(l)\}_{m=0,1,2,\cdots}$ form POVMs

$$\sum_{i=g,e} A_i^\dagger(l) A_i(l) = I_{\text{photon}}, \tag{41}$$

$$\sum_{m=0}^{\infty} B_m^\dagger(l) B_m(l) = I_{\text{atom}}, \tag{42}$$

with $I$'s being the identity operators. Using Eqs. (5) and (36), they are explicitly calculated as follows:

$$A_i(l) = \langle i|U^\dagger(l)|e\rangle = \sum_{n=0}^{\infty} \left[\alpha_n(l)\, \delta_{i,e}|n\rangle\langle n| + \gamma_n(l)\delta_{i,g}|n+1\rangle\langle n|\right], \tag{43}$$

$$B_m(l) = \langle m|U^\dagger(l)|n\rangle$$

$$= \delta_{n-1}(l)\sigma_+\, \delta_{m,n-1} + [\alpha_n(l)\sigma_+\sigma_- + \beta_{n-1}(l)\sigma_-\sigma_+]\,\delta_{m,n} + \gamma_n(l)\sigma_-\delta_{m,n+1}. \tag{44}$$

These operators do not satisfy the unital conditions: $\sum_{i=g,e} A_i(l)\, A_i^\dagger(l) \neq I_{\text{photon}}$,

$\sum_{m=0}^{\infty} B_m(l)\, B_m^\dagger(l) \neq I_{\text{atom}}$. Accordingly, entropy, which will be discussed in the next section, does not increase under the operations of these POVMs [12,13], in general: that is, whether entropy increases by the operations depends on choice of a state to be transformed.

## V. STATIONARY PHOTON-ATOM ENTANGLEMENT

As mentioned in the Introduction, the flow equation approach may offer the optimal strategy for generating the stationary state of the Hamiltonian. This implies that it is possible to measure how quantum-state engineering makes a given "initial" state close to the stationary state.

As mentioned in the previous section, taking into account a realistic experimental situation using a cavity, a natural "initial" state is a product state. Since $U(l)$ in Eq. (5) is a nonlocal unitary operator, it necessarily creates entanglement between the photons and atom. Then, it is of interest to quantify the degree of entanglement in the engineered state as well as its closeness to the exact stationary state. Here, we discuss this issue by analyzing the entanglement entropies.

The entanglement entropies are defined by the von Neumann entropies of the subsystems:

$$S_{\text{photon}}(l) = -\mathrm{Tr}_{\text{photon}}\, \rho_{\text{photon}}(l) \ln \rho_{\text{photon}}(l), \tag{45}$$

$$S_{\text{atom}}(l) = -\text{Tr}_{\text{atom}}\, \rho_{\text{atom}}(l) \ln \rho_{\text{atom}}(l), \tag{46}$$

where $\rho_{\text{photon}}(l)$ and $\rho_{\text{atom}}(l)$ are the reduced density matrices defined in the previous section. These are nonzero because $|\Psi(l)\rangle$ in eq. (37) is an entangled state for $l > 0$. From the Araki-Lieb inequalities [14] in the present case,

$$\left| S_{\text{photon}}(l) - S_{\text{atom}}(l) \right| \le S_{\text{total}}(l) \le S_{\text{photon}}(l) + S_{\text{atom}}(l), \tag{47}$$

it follows that the entanglement entropies of the photons and atom are identical

$$S_{\text{photon}}(l) = S_{\text{atom}}(l) \tag{48}$$

at each value of *l*, since $S_{\text{total}}(l)$ vanishes because of purity of the total state in Eq. (38). Therefore, henceforth we consider only $S_{\text{atom}}(l)$.

An explicit calculation yields

$$S_{\text{atom}}(l) = -s_+(l) \ln s_+(l) - s_-(l) \ln s_-(l), \tag{49}$$

$$s_\pm(l) = \frac{1 \pm \tilde{L}_n(l)}{2} \quad \left( \xrightarrow{l \to \infty} \frac{1}{2}\left[1 \pm \frac{\Delta}{\Omega_n}\right] \right), \tag{50}$$

where $\tilde{L}_n(l)$ is given in Eq. (30). Numerical analysis of $S_{\rm atom}(l)$ with the use of the physical parameters taken from real experiments (for example, as in Refs. [15,16]) shows a monotonic behavior, exhibiting how the degree of entanglement increases as the "initial" product state approaches the exact stationary state. The flow of $S_{\rm atom}(l)$ can be also used for quantifying the deviation of an engineered state from the stationary state.

Since $|\Psi(l)\rangle$ with a finite value of $l$ is not the stationary state, its time evolution makes the entanglement entropies oscillatory. To see this, we have calculated the time evolution of $S_{\rm atom}(l,t)$ for $|\Psi(l,t)\rangle \equiv e^{-iHt}|\Psi(l)\rangle$. The result is

$$S_{\rm atom}(l,t) = -s_+(l,t)\ln s_+(l,t) - s_-(l,t)\ln s_-(l,t), \tag{51}$$

$$s_\pm(l,t) = \frac{1}{2}\Bigg\{ 1 \pm \tilde{L}_n(l)\cos\Omega_n t \pm (1-\cos\Omega_n t)\frac{\Delta}{\Omega_n}\left[\frac{\Delta}{\Omega_n}\tilde{L}_n(l) + \sqrt{\left(1-\tilde{L}_n^2(l)\right)\left(1-\frac{\Delta^2}{\Omega_n^2}\right)}\right]\Bigg\}. \tag{52}$$

Numerical analysis of $S_{\rm atom}(l,t)$ may show how the oscillation decays as $l$ increases.

## VI. COMMENT ON IMPLEMENTAION OF THE UNITARY OPERATION

Finally, we discuss how to approximately implement the unitary operation appearing in the method of flow equation for the Jaynes-Cummings model.

In a real experimental condition [15], $\lambda / \Delta = O(10^{-3})$. Therefore, expanding the coefficients, $\alpha$'s, $\beta$'s, $\gamma$'s, and $\delta$'s in Eq. (5) with respect to it, we obtain the following approximate expression:

$$U(l) \cong I_{\text{photon}} \otimes I_{\text{atom}} + \left(1 - e^{-\Delta^2 l}\right) \frac{\lambda}{\Delta} (\sigma_+ a - \sigma_- a^\dagger). \tag{53}$$

The above operator is linear in the photon field, and therefore it may not be difficult to be implemented in terms of the quadrature operators. On the other hand, for the atom operators, we note that the use of the Ramsey zone [17] yields the transformation, $|e\rangle \to \left(|e\rangle + |g\rangle\right) / \sqrt{2}$, which can be generated by the following unitary operator:

$$R_{\theta,\phi} = \frac{1}{\sqrt{2}} \left( \sigma_+ \sigma_- + e^{i\theta} \sigma_- \sigma_+ + e^{i\phi} \sigma_+ + \sigma_- \right), \tag{54}$$

where the phases satisfy the condition, $\theta - \phi = \pi \bmod (2\pi)$. Therefore, we find the atom operators to be given as follows:

$$\sigma_+ = \frac{1}{2}\left[\sqrt{2}\left(R_{\pi,0} - R_{0,\pi}\right) - R_{\pi,0}\,R_{0,\pi} + I_{\text{atom}}\right], \tag{55}$$

$$\sigma_- = \frac{1}{2}\left[\sqrt{2}\left(R_{\pi,0} + R_{0,\pi}\right) - R_{\pi,0}\,R_{0,\pi} - I_{\text{atom}}\right]. \tag{56}$$

Therefore, the unitary transformation of the flow equation can approximately be implemented by making use of the quadrature operators of the photon field and the Ramsey zone.

## VII. CONCLUSION

We have discussed stationary photon-atom entanglement in the Jaynes-Cummings model by applying the method of flow equation. We have explicitly constructed the nonlocal continuous unitary operator and the associated positive operator-valued measures for the photons and atom as the subsystems. We have analyzed flow of the entanglement entropies and furthermore have discussed an approximate implementation of the unitary operation.

As emphasized in the Introduction, it is our opinion that Wegner’s flow equation is more than just a mathematical tool for diagonalizing a given Hamiltonian. In the present work, we have shown how it offers a useful approach for quantum-state engineering. We believe that this method may be able to play a peculiar role in quantum information technology.

## ACKNOWLEDGMENTS

Y. I. would like to thank H. Hatano at Ritsumeikan University for discussions about quantum optical experiments using two-level atoms. S. A. was supported in part by Grant-in-Aid for Scientific Research (B) of the Ministry of Education.

---